\documentclass[a4paper,12pt]{article}
\usepackage{amsthm,amsmath,amssymb,latexsym}

\newtheorem{thm}{Theorem}[section]
\newtheorem{prop}[thm]{Proposition}
\newtheorem{lem}[thm]{Lemma}
\newtheorem{rem}[thm]{Remark}
\numberwithin{equation}{section}

\newcommand{\ds}{\displaystyle}
\newcommand{\vep}{\varepsilon}
\newcommand{\wt}{\widetilde}
\newcommand{\wh}{\widehat}
\newcommand{\DynB}{
\begin{figure}
\begin{center}
\begin{picture}(165,70)(10,-10)
\put(20,0){\circle{6}}\put(22,2){\line(1,1){16}}
\put(20,40){\circle{6}}\put(22,38){\line(1,-1){16}}
\put(40,20){\circle{6}}\put(43,20){\line(1,0){24}}
\put(70,20){\circle{6}}\put(73,20){\line(1,0){14}}
\multiput(89,20)(3,0){8}{\line(1,0){1}}\put(113,20){\line(1,0){14}}
\put(130,20){\circle{6}}\put(132,22){\vector(1,0){26}}
\put(132,18){\vector(1,0){26}}
\put(160,20){\circle{6}}
\put(24,42){\small$\alpha_0$}
\put(18,-8){\small$\alpha_1$}
\put(40,12){\small$\alpha_2$}
\put(68,12){\small$\alpha_3$}
\put(128,12){\small$\alpha_{n+2}$}
\put(158,12){\small$\alpha_{n+3}$}
\end{picture}
\end{center}
\caption{Dynkin diagram of type $B^{(1)}_{n+3}$}
\end{figure}
}
\newcommand{\DynF}{
\begin{figure}
\begin{center}
\begin{picture}(165,30)(10,-10)
\put(30,10){\circle{6}}\put(33,10){\line(1,0){24}}
\put(60,10){\circle{6}}\put(63,10){\line(1,0){24}}
\put(90,10){\circle{6}}
\put(92,12){\vector(1,0){26}}\put(92,8){\vector(1,0){26}}
\put(120,10){\circle{6}}\put(123,10){\line(1,0){24}}
\put(150,10){\circle{6}}
\put(30,2){\small$\alpha_0$}
\put(60,2){\small$\alpha_1$}
\put(90,2){\small$\alpha_2$}
\put(120,2){\small$\alpha_3$}
\put(150,2){\small$\alpha_4$}
\end{picture}
\end{center}
\caption{Dynkin diagram of type $F^{(1)}_{4}$}
\end{figure}
}
\DeclareMathOperator*{\res}{Res}

\title{Affine Weyl group symmetry of the Garnier system}
\author{Takao Suzuki\\
Department of Mathematics, Kobe University,\\
Rokko, Kobe 657-8501, Japan\\
{\small suzukit@math.kobe-u.ac.jp}}
\date{}

\begin{document}

\maketitle

\begin{abstract}
In this paper, we show that the Garnier system in $n$-variables has affine Weyl
group symmetry of type $B^{(1)}_{n+3}$.
We also formulate the $\tau$-functions for the Garnier system (or the
Schlesinger system of rank 2) on the root lattice $Q(C_{n+3})$ and show that
they satisfy Toda equations, Hirota-Miwa equations and bilinear differential
equations.
\end{abstract}

\section{Introduction}
For the sixth Painlev\'{e} equation $P_{VI}$, the symmetry structure is
well-known \cite{IKSY,OKM1}.
Furthermore, the $\tau$-functions for $P_{VI}$ satisfy various bilinear
relations \cite{MSD,OKM1,OKM2}.
But such properties are not clarified completely for the Garnier system which
is an extension of $P_{VI}$ to several variables.
In this paper, we show that the Garnier system in $n$-variables ($n\geq2$) has
affine Weyl group symmetry of type $B^{(1)}_{n+3}$.
We also formulate the $\tau$-functions for the Garnier system (or the
Schlesinger system of rank 2) on the root lattice $Q(C_{n+3})$ and show that
they satisfy Toda equations, Hirota-Miwa equations and bilinear differential
equations.

Consider a Fuchsian differential equation on $\mathbb{P}^1(\mathbb{C})$
\begin{equation}\label{Eq:Fuchs}
    \frac{d^2y}{dz^2}+P_1(z)\,\frac{dy}{dz}+P_2(z)\,y = 0
\end{equation}
with regular singularities $z=t_1,\ldots,t_n$, $t_{n+1}=0$, $t_{n+2}=1$,
$\infty$, apparent singularities $z=\lambda_1,\ldots,\lambda_n$ and the Riemann
scheme
\begin{equation}\label{Eq:Rie_Scm}
	\left(\begin{array}{ccccccc}
		z=t_1& \ldots& z=t_{n+2}& z=\infty& z=\lambda_1& \ldots& z=\lambda_n\\
		0& \ldots& 0& \rho& 0& \ldots& 0\\
		\theta_1& \ldots& \theta_{n+2}& \rho+\theta_{n+3}+1& 2& \ldots& 2
	\end{array}\right),
\end{equation}
assuming that the Fuchs relation
\begin{equation}
	\sum_{j=1}^{n+3}\,\theta_j+2\rho = 1
\end{equation}
is satisfied.
The monodromy preserving deformations of the equation \eqref{Eq:Fuchs} with the
scheme \eqref{Eq:Rie_Scm} is described as the following completely integrable
Hamiltonian system \cite{IKSY}:
\begin{equation}\label{Eq:Gar_Rat}
	\frac{\partial\lambda_j}{\partial t_i}
	= \frac{\partial\mathcal{K}_i}{\partial\mu_j},\quad
	\frac{\partial\mu_j}{\partial t_i}
	= -\frac{\partial\mathcal{K}_i}{\partial\lambda_j}\quad (i,j=1,\ldots,n),
\end{equation}
where
\begin{equation}
	\mu_j = \res_{z=\lambda_j}P_2(z)\,dz\quad (j=1,\ldots,n)
\end{equation}
and $\mathcal{K}_i$ $(i=1,\ldots,n)$ are rational functions in $\lambda_j$,
$\mu_j$ $(j=1,\ldots,n)$ given by
\begin{equation}
	\mathcal{K}_i = -\res_{z=t_i}P_2(z)\,dz.
\end{equation}
By the canonical transformation
\begin{equation}
	x_i = \frac{t_i}{t_i-1},\quad
	q_i = \frac{t_i\prod_{j=1}^{n}(t_i-\lambda_j)}
	{\prod_{j=1,j\neq i}^{n+2}(t_i-t_j)}\quad (i=1,\ldots,n),
\end{equation}
the system \eqref{Eq:Gar_Rat} is transformed into the Hamiltonian system
\begin{equation}\label{Eq:Sys_Gar}
	\frac{\partial q_j}{\partial x_i}
	= \frac{\partial K_i}{\partial p_j},\quad
	\frac{\partial p_j}{\partial x_i}
	= -\frac{\partial K_i}{\partial q_j}\quad (i,j=1,\ldots,n)
\end{equation}
with {\it polynomial} Hamiltonians $K_i$ $(i=1,\ldots,n)$.
These $K_i$ are given explicitly by
\begin{equation}\label{Eq:Ham_Gar}
	\begin{split}
		x_i(x_i-1)K_i &= q_i\left(\rho+\sum_{j=1}^{n}q_jp_j\right)
		\left(\rho+\theta_{n+3}+1+\sum_{j=1}^{n}q_jp_j\right)
		+ x_ip_i(q_ip_i-\theta_i)\\
		&\quad
		- \sum_{j=1,j\neq i}^{n}X_{ij}\,q_ip_i(q_jp_j-\theta_j)
		- \sum_{j=1,j\neq i}^{n}X_{ji}\,q_i(q_jp_j-\theta_j)p_j\\
		&\quad
		- \sum_{j=1,j\neq i}^{n}X^*_{ij}\,(q_ip_i-\theta_i)p_iq_j
		- \sum_{j=1,j\neq i}^{n}X_{ij}\,(q_ip_i-\theta_i)q_jp_j\\
		&\quad
		- (x_i+1)(q_ip_i-\theta_i)q_ip_i
		+ (\theta_{n+2}\,x_i+\theta_{n+1}-1)q_ip_i,
	\end{split}
\end{equation}
where
\begin{equation}
	X_{ij} = \frac{x_i(x_j-1)}{x_j-x_i},\quad
	X^*_{ij} = \frac{x_i(x_i-1)}{x_i-x_j}.
\end{equation}
We call the Hamiltonian system \eqref{Eq:Sys_Gar} with the Hamiltonians
\eqref{Eq:Ham_Gar} the {\it Garnier system}.

As is known in \cite{IKSY}, the Garnier system is derived from the Schlesinger
system (of rank 2).
Then the independent and dependent variables of the Garnier system are
expressed as certain rational functions in the variables of the Schlesinger
system.
Furthermore, the $\tau$-functions for the Garnier system can be identified with 
those for the Schlesinger system.
Hence we first investigate symmetries and properties of the $\tau$-functions
for the Schlesinger system.
After that, we apply the obtained results to the Garnier system.

In Section \ref{Sec:Sch}, we give the transformations of three types, permutation of the points, sign change of the exponents and Schlesinger
transformation, which act on the Schlesinger system.
In Section \ref{Sec:Tau_Sch}, we formulate the $\tau$-functions for the
Schlesinger system on the root lattice $Q(C_{n+3})$.
We also present bilinear relations which are satisfied by the $\tau$-functions.
In Section \ref{Sec:Gar}, we show that the Garnier system has affine Weyl group
symmetry of type $B^{(1)}_{n+3}$.

\section{Schlesinger system}\label{Sec:Sch}
Let $A_j$ and $G_j$ $(j=1,\ldots,n+2)$ be matrices of dependent variables
defined as
\begin{equation}
	A_j = \begin{pmatrix}a_j&b_j\\c_j&d_j\end{pmatrix},\quad
	G_j = \begin{pmatrix}-d_j&b_j\\c_j&d_j\end{pmatrix}
	\begin{pmatrix}g_j&0\\0&h_j\end{pmatrix}.
\end{equation}
Consider a system of total differential equations
\begin{equation}\label{Sys:Schlesinger}
	\begin{array}{ll}
		\ds dA_j = \sum_{i=1,i\neq j}^{n+2}[A_i,A_j]\,d\log\,(t_j-t_i)&
		(j=1,\ldots,n+2),\\[16pt]
		\ds dG_j = \sum_{i=1,i\neq j}^{n+2}A_iG_j\,d\log\,(t_j-t_i)&
		(j=1,\ldots,n+2),
    \end{array}
\end{equation}
where $t_{n+1}=0$, $t_{n+2}=1$ and $d$ is an exterior differentiation with
respect to $t_1,\ldots,t_n$.
Here we assume
\begin{enumerate}
\item[(i)]\quad
$\mathrm{det}A_j=0$,\quad $\mathrm{tr}A_j=\theta_j\notin\mathbb{Z}$\quad
$(j=1,\ldots,n+2)$;
\item[(ii)]\quad
$-\sum_{j=1}^{n+2}A_j=\mathrm{diag}\,(\rho,\,\rho+\theta_{N+3})$,\quad
$\theta_{n+3}\notin\mathbb{Z}$,\quad $\rho = -\sum_{j=1}^{n+3}\theta_j/2$.
\end{enumerate}
We call the system \eqref{Sys:Schlesinger} the {\it Schlesinger system}.

Recall that the Schlesinger system is obtained as the compatibility condition
for a system of linear differential equations on $\mathbb{P}^{1}(\mathbb{C})$
\begin{equation}\label{Eq:Lax_Sch}
	\frac{\partial\vec{y}}{\partial z}
	= \sum_{j=1}^{n+2}\frac{A_j}{z-t_j}\,\vec{y},\quad
	\frac{\partial\vec{y}}{\partial t_i}
	= -\frac{A_i}{z-t_i}\,\vec{y}\quad (i=1,\ldots,n),
\end{equation}
where $\vec{y}={}^t(y_1,y_2)$ is a vector of unknown functions.
The matrices $G_j$ $(j=1,\ldots,n+2)$ are obtained as follows.
The system \eqref{Eq:Lax_Sch} has a local fundamental solution $Y=Y(z)$ of the
form
\begin{equation}
	Y = Y_j(z)\,(z-t_j)^{\theta_jE_2}\quad (j=1,\ldots,n+2),
\end{equation}
where
\begin{equation}
	E_1 = \begin{pmatrix}1&0\\0&0\end{pmatrix},\quad
	E_2 = \begin{pmatrix}0&0\\0&1\end{pmatrix}.
\end{equation}
Here $Y_j(z)$ is a $2\times2$ matrix which is holomorphic at $z=t_j$, such
that
\begin{equation}
	Y_j(z)\bigm|_{z=t_j} = G_j,\quad G_j^{-1}A_jG_j = \theta_jE_2.
\end{equation}

Note that the Schlesinger system has an ambiguity for the following
transformation:
\begin{equation}
	A_j\to C^{-1}A_jC,\quad G_j\to C^{-1}G_j\quad (j=1,\ldots,n+2),
\end{equation}
where
\begin{equation}
	C = \begin{pmatrix}\gamma_1&0\\0&\gamma_2\end{pmatrix}\quad
	(\gamma_1,\gamma_2\in\mathbb{C}).
\end{equation}

The Schlesinger system is invariant under the action of the following
transformations of three types.
They are associated with (1) permutation of the points $t_1,\ldots,t_{n+2}$,
$t_{n+3}=\infty$, (2) sign change of the exponents
$\theta_1,\ldots,\theta_{n+3}$, and (3) shifting of the exponents by integers
({\it Schlesinger transformation}).
In this section, we describe these transformations.

\subsection{Permutation of the points}
In the following, we use the matrix notations
\begin{equation}
	w(A_j) = \begin{pmatrix}w(a_j)&w(b_j)\\w(c_j)&w(d_j)\end{pmatrix}
\end{equation}
and
\begin{equation}
	w(G_j)
	= \begin{pmatrix}-w(d_j)&w(b_j)\\w(c_j)&w(d_j)\end{pmatrix}
	\begin{pmatrix}w(g_j)&0\\0&w(h_j)\end{pmatrix}
\end{equation}
for a transformation $w$ of the dependent variables.

The action of the symmetric group $\mathfrak{S}_{n+3}$ on the set of the points
$t_1,\ldots,t_n$, $t_{n+1}=0$, $t_{n+2}=1$, $t_{n+3}=\infty$ can be lifted to
transformations of the independent and dependent variables.
Denoting the adjacent transpositions by
$\sigma_1=(12),\ldots,\sigma_{n+2}=(n+2,n+3)$, we describe the action of these
$\sigma_k$ on the variables $t_i$ $(i=1,\ldots,n)$ and $a_j$, $b_j$, $c_j$,
$d_j$, $g_j$, $h_j$ $(j=1,\ldots,n+2)$.
\begin{equation}
	\sigma_k(t_i) = t_{\sigma_k(i)},\quad
	\sigma_k(A_j) = A_{\sigma_k(j)},\quad
	\sigma_i(G_j) = G_{\sigma_i(j)}
\end{equation}
for $k=1,\ldots,n-1$.
We remark that $\sigma_n$, $\sigma_{n+1}$ and $\sigma_{n+2}$ are derived from
M\"{e}bius transformations on $\mathbb{P}^1(\mathbb{C})$.
The transformation $\sigma_n$ is derived from $z\to(z-t_n)\,/\,(1-t_n)$:
\begin{equation}
	\begin{split}
		&\sigma_n(t_i) = \frac{t_i-t_n}{1-t_n}\quad (i\neq n),\quad
		\sigma_n(t_n) = \frac{-t_n}{1-t_n},\\
		&\sigma_n(A_j) = (1-t_n)^{\theta_{n+3}E_2}A_{\sigma_n(j)}\,
		(1-t_n)^{-\theta_{n+3}E_2},\\
		&\sigma_n(G_j) = (1-t_n)^{\rho I_2+\theta_{n+3}E_2}G_{\sigma_n(j)}\,
		(1-t_n)^{\theta_{\sigma_n(j)}E_2}.
	\end{split}
\end{equation}
Similarly, the transformation $\sigma_{n+1}$ is derived from $z\to1-z$:
\begin{equation}
	\sigma_{n+1}(t_i) = 1-t_i,\quad
	\sigma_{n+1}(A_j) = A_{\sigma_{n+1}(j)},\quad
	\sigma_{n+1}(G_j)=G_{\sigma_{n+1}(j)},
\end{equation}
and the transformation $\sigma_{n+2}$ is derived from $z\to1/z$:
\begin{equation}
	\begin{array}{ll}
		\ds\sigma_{n+2}(t_i) = \frac{t_i}{t_i-1},&\\[8pt]
		\sigma_{n+2}(A_j) = G_{n+2}^{-1}A_jG_{n+2}& (j\neq n+2),\\[4pt]
		\sigma_{n+2}(A_{n+2}) = \theta_{n+3}\,G_{n+2}^{-1}E_2G_{n+2},\\[4pt]
		\sigma_{n+2}(G_j) = G_{n+2}^{-1}G_j\,(t_j-1)^{\rho I_2+2\theta_jE_2}&
		(j\neq n+2),\\[4pt]
		\sigma_{n+2}(G_{n+2}) = G_{n+2}^{-1},&
	\end{array}
\end{equation}
The action of each $\sigma_k$ on the parameters $\theta_j$ is given by
\begin{equation}
	\sigma_k(\theta_j) = \theta_{\sigma_k(j)}\quad (j=1,\ldots,n+3).
\end{equation}

\subsection{Sign change of the exponents}
Let $Y$ be a fundamental solution of system \eqref{Eq:Lax_Sch}.
Consider the gauge transformations
\begin{equation}
	r_k(Y) = (z-t_k)^{-\theta_k}Y\quad (k=1,\ldots,n+2),\quad
	r_{n+3}(Y) = WY,
\end{equation}
where
\begin{equation}
	W = \begin{pmatrix}0&1\\1&0\end{pmatrix}.
\end{equation}
Each $r_k$ acts on the parameters as follows:
\begin{equation}
	r_k(\theta_j) = (-1)^{\delta_{jk}}\theta_j\quad (j=1,\ldots,n+3),
\end{equation}
where $\delta_{jk}$ stands for the Kronecker delta, and can be lifted to a
transformation of the dependent variables.
We describe the action of these $r_k$.
\begin{equation}
	\begin{array}{ll}
		r_k(A_j) = A_j-\delta_{jk}\,\theta_kI_2& (j=1,\ldots,n+2),\\[4pt]
		r_k(G_j) = (t_j-t_k)^{-\delta_{jk}\theta_k}G_j& (j=1,\ldots,n+2)
	\end{array}
\end{equation}
for $k=1,\ldots,n+2$.
\begin{equation}
	r_{n+3}(A_j) = WA_jW,\quad r_{n+3}(G_j) = WG_j\quad (j=1,\ldots.n+2)
\end{equation}
for $k=n+3$.
Note that the independent variables $t_i$ $(i=1,\ldots,n)$ are invariant under
the action of each $r_k$.

\subsection{Schlesinger transformations}\label{Subsec:Sch_Trf}
In this section, we construct the Schlesinger transformations by following
\cite{JM}.
Let $L$ be a subset of $\mathbb{Z}^{n+3}$ defined as
\begin{equation}
	L = \left\{\mu=(\mu_1,\ldots,\mu_{n+3})\in\mathbb{Z}^{n+3}\bigm|
	\mu_1+\ldots+\mu_{n+3}\in2\,\mathbb{Z}\right\}.
\end{equation}
Consider the gauge transformations
\begin{equation}
    T_{\mu}(Y) = R_{\mu}Y\quad (\mu\in L),
\end{equation}
where $R_{\mu}$ are $2\times2$ matrices of rational functions in $z$ and $t_i$
$(i=1,\ldots,n)$, such that
\begin{equation}
	T_{\mu}(\theta_j) = \theta_j+\mu_j\quad (j=1,\ldots,n+3).
\end{equation}
Then each $R_{\mu}$ is determined up to multiplication by a scalar matrix and
the gauge transformation $T_{\mu}$ can be lifted to a birational transformation
(called the Schlesinger transformation) of the dependent variables.

The group of the Schlesinger transformations is generated by the
transformations $T_k$ $(k=1,\ldots,n+2)$, such that
\begin{equation}
	T_k(\theta_j) = \theta_j+\delta_{jk}-\delta_{j\,k+1}\quad (j=1\ldots,n+3),
\end{equation}
and $T_{n+3}$, such that
\begin{equation}
	T_{n+3}(\theta_j) = \theta_j+\delta_{j\,n+2}+\delta_{j\,n+3}\quad
	(j=1\ldots,n+3).
\end{equation}
We describe the action of these $T_k$ on the variables $a_j$, $b_j$, $c_j$,
$d_j$, $g_j$, $h_j$ $(j=1,\ldots.n+2)$.
\begin{equation}
	\begin{split}
		T_k(A_j) &= A_j + \frac{R_k^*A_jR_k}{(t_k-t_{k+1})(t_k-t_j)}
		- \frac{R_kA_jR_k^*}{(t_k-t_{k+1})(t_{k+1}-t_j)}\quad (j\neq k,k+1),\\
		T_k(A_k) &= A_{k+1} - \frac{(1+\theta_k-\theta_{k+1})R_k}{t_k-t_{k+1}}
		- \sum_{j=1,j\neq k,k+1}^{n+2}
		\frac{R_k^*A_jR_k}{(t_k-t_{k+1})(t_k-t_j)},\\
		T_k(A_{k+1}) &= A_k + \frac{(1+\theta_k-\theta_{k+1})R_k}{t_k-t_{k+1}}
		+ \sum_{j=1,j\neq k,k+1}^{n+2}
		\frac{R_kA_jR_k^*}{(t_k-t_{k+1})(t_{k+1}-t_j)},\\
		T_k(G_j) &= G_j - \frac{R_kG_j}{t_{k+1}-t_j}\quad (j\neq k,k+1),\\
		T_k(G_k) &= \frac{R_k^*G_k}{t_k-t_{k+1}} + \frac{G_kE_2}{t_k-t_{k+1}}
		+ \sum_{j=1,j\neq k}^{n+2}\frac{R_k^*G_kE_1G_k^{-1}A_jG_kE_2}
		{(1+\theta_k)(t_k-t_{k+1})(t_k-t_j)},\\
		T_k(G_{k+1}) &= R_kG_{k+1} - G_{k+1}E_2
		+ \sum_{j=1,j\neq k}^{n+2}
		\frac{R_kG_{k+1}E_2G_k^{-1}A_jG_kE_1}{(1-\theta_k)(t_k-t_j)},
	\end{split}
\end{equation}
where
\begin{equation}
	R_k = \frac{-t_k+t_{k+1}}{b_ka_{k+1}+d_kb_{k+1}}
	\left(\begin{array}{@{}c@{}}b_{k}\\d_{k}\end{array}\right)
	\left(\begin{array}{@{}cc@{}}a_{k+1}&b_{k+1}\end{array}\right),\quad
	R_k^* = (t_k-t_{k+1})I_2 + R_k,
\end{equation}
for $k=1,\ldots,n+1$.
\begin{equation}
	\begin{split}
		T_{n+2}(A_{n+2}) &= R_{n+2}A_{n+2}E_1 + E_2A_{n+2}R_{n+2}^*
		+ E_2R_{n+2}^* - \sum_{j=1}^{n+1}\,\frac{R_{n+2}A_jR_{n+2}^*}{t_j-1},\\
		T_{n+2}(A_j) &= (t_j-1)E_2A_jE_1 + R_{n+2}A_jE_1 + E_2A_jR_{n+2}^*
		+ \frac{R_{n+2}A_jR_{n+2}^*}{t_j-1}\\
		&\mspace{360mu} (j\neq n+2),\\
		T_{n+2}(G_{n+2}) &= R_{n+2}G_{n+2} + E_2G_{n+2}E_2
		+ \sum_{j=1,j\neq k}^{n+2}
		\frac{R_{n+2}G_{n+2}E_1G_{n+2}^{-1}A_jG_{n+2}E_2}
		{(1+\theta_{n+2})(t_{n+2}-t_j)},\\
		T_{n+2}(G_j)&= (t_j-1)E_2G_j+R_{n+2}G_j\quad (j\neq n+2),
	\end{split}
\end{equation}
where
\begin{equation}
	\begin{split}
		&R_{n+2} = \frac{1}{(1-\theta_{n+3})\,d_{n+2}}
		\left(\begin{array}{@{}c@{}}1-\theta_{n+3}\\c_{\infty}\end{array}
		\right)
		\left(\begin{array}{@{}cc@{}}d_{n+2}&-b_{n+2}\end{array}\right),\\
		&R_{n+2}^* = \frac{1}{(1-\theta_{n+3})\,d_{n+2}}
		\left(\begin{array}{@{}c@{}}b_{n+2}\\d_{n+2}\end{array}\right)
		\left(\begin{array}{@{}cc@{}}-c_{\infty}&1-\theta_{n+3}\end{array}
		\right)
	\end{split}
\end{equation}
and $c_{\infty}=\sum_{j=1}^{n+2}t_jc_j$, for $k=n+2$.
\begin{equation}
	\begin{split}
		T_{n+3}(A_{n+2}) &= R_{n+3}A_{n+2}E_2 + E_1A_{n+2}R_{n+3}^*
		+ E_1R_{n+3}^* - \sum_{j=1}^{n+1}\,\frac{R_{n+3}A_jR_{n+3}^*}{t_j-1},\\
		T_{n+3}(A_j) &= (t_j-1)E_1A_jE_2 + R_{n+3}A_jE_2 + E_1A_jR_{n+3}^*
		+ \frac{R_{n+3}A_jR_{n+3}^*}{t_j-1}\\
		&\mspace{360mu} (j\neq n+2),\\
		T_{n+3}(G_{n+2}) &= R_{n+3}G_{n+2} + E_1G_{n+2}E_2
		+ \sum_{j=1,j\neq k}^{n+2}
		\frac{R_{n+3}G_{n+2}E_1G_{n+2}^{-1}A_jG_{n+2}E_2}
		{(1+\theta_{n+2})(t_{n+2}-t_j)},\\
		T_{n+3}(G_j) &= (t_j-1)E_1G_j+R_{n+3}G_j\quad (j\neq n+2),
	\end{split}
\end{equation}
where
\begin{equation}
	\begin{split}
		&R_{n+3} = \frac{1}{(1+\theta_{n+3})\,b_{n+2}}
		\left(\begin{array}{@{}c@{}}b_{\infty}\\1+\theta_{n+3}\end{array}
		\right)
		\left(\begin{array}{@{}cc@{}}-d_{n+2}&b_{n+2}\end{array}\right),\\
		&R_{n+3}^* = \frac{1}{(1+\theta_{n+3})\,b_{n+2}}
		\left(\begin{array}{@{}c@{}}b_{n+2}\\d_{n+2}\end{array}\right)
		\left(\begin{array}{@{}cc@{}}1+\theta_{n+3}&-b_{\infty}\end{array}
		\right)
	\end{split}
\end{equation}
and $b_{\infty}=\sum_{j=1}^{n+2}t_jb_j$, for $k=n+3$.
Note that the independent variables $t_i$ $(i=1,\ldots,n)$ are invariant under
the action of each $T_k$.

\begin{rem}
The group of the Schlesinger transformations generated by $T_k$
$(k=1,\ldots,n+3)$ is isomorphic to the root lattice $Q(C_{n+3})$.
The commutativity between two arbitrary Schlesinger transformations is obtained
from the uniqueness of the Schlesinger transformations {\rm\cite{JM}}.
\end{rem}

\section{$\tau$-Functions on the root lattice}\label{Sec:Tau_Sch}
In this Section, we formulate the $\tau$-functions for the Schlesinger system
on the root lattice $Q(C_{n+3})$.
We also present the bilinear relations of three types, Toda equations,
Hirota-Miwa equations and bilinear differential equations, which are satisfied
by the $\tau$-functions.

\begin{prop}[\rm\cite{JM}\bf]\label{Prop:1-form_Sch}
For each solution of the Schlesinger system, the 1-forms
\begin{equation}
	\omega_{\mu} = \sum_{i=1}^{N}\,T_{\mu}(H_i)\,dt_i\quad (\mu\in L)
\end{equation}
are closed.
Here we let
\begin{equation}\label{Eq:Ham_Sch}
	H_i = \sum_{j=1,j\neq i}^{n+2}
	\frac{1}{t_i-t_j}\left(\mathrm{tr}A_iA_j+C_{ij}\right)\quad
	(i=1,\ldots,n),
\end{equation}
where
\begin{equation}
	C_{ij} = -\frac{1}{2}\,\theta_i\theta_j
	+ \frac{\theta_i^2+\theta_j^2}{2\,(n+1)}
	- \frac{\sum_{i=1}^{n+3}\theta_i^2}{2\,(n+1)(n+2)}.
\end{equation}
\end{prop}
Proposition \ref{Prop:1-form_Sch} allows us to define a family of
$\tau$-functions by
\begin{equation}
	d\log\tau_{\mu} = \omega_{\mu}\quad (\mu\in L),
\end{equation}
up to multiplicative constants.

We also define the action of the transformations $\sigma_k$, $r_l$ and
$T_{\mu}$ on the $\tau$-functions, so that it is consistent with the action of
them on $H_i$ which we call Hamiltonians.
For each $\mu,\nu\in L$, the action of $T_{\mu}$ on $\tau_{\nu}$ is defined by
\begin{equation}
	T_{\mu}(\tau_{\nu}) = \tau_{\mu+\nu}
\end{equation}
and the action of $\sigma_k$, $r_l$ on $\tau_{\nu}$ is defined by
\begin{equation}
	\begin{array}{ll}
		\sigma_k(\tau_{\nu}) = \tau_{\sigma_k(\nu)}& (k=1,\ldots,n+2),\\[4pt]
		r_l(\tau_{\nu}) = \tau_{r_l(\nu)}& (l=1,\ldots,n+3),
	\end{array}
\end{equation}
where
\begin{equation}
	\begin{split}
		\sigma_k(\nu)
		&= (\nu_{\sigma_k(1)},\ldots,\nu_{\sigma_k(n+3)}),\\
		r_l(\nu)
		&= (\nu_1,\ldots,\nu_{l-1},-\nu_l,\nu_{l+1},\ldots,\nu_{n+3}).
	\end{split}
\end{equation}

In Section \ref{SubSec:Sym_Ham}, we describe the action of the transformations
$\sigma_k$, $r_l$ and $T_{\mu}$ on the Hamiltonians, which is obtained
from the action of them on the independent and dependent variables.

\subsection{Symmetries for Hamiltonians}\label{SubSec:Sym_Ham}
We first describe the action of the Schlesinger transformation $T_{\mu}$ on the
Hamiltonians for each $\mu\in L$ with
\begin{equation}
	\mu_1^2 + \ldots + \mu_{n+3}^2 = 2.
\end{equation}
Set
\begin{equation}
	\begin{split}
		T_{k,l} = T_{\mathbf{e}_k+\mathbf{e}_l},\quad
		T_{k,-l} = T_{\mathbf{e}_k-\mathbf{e}_l},\quad
		T_{-k,-l} = T_{-\mathbf{e}_k-\mathbf{e}_l}\\
		(k,l=1,\ldots,n+3,\,k\neq l),
	\end{split}
\end{equation}
where
\begin{equation}
	\begin{split}
		\mathbf{e}_1 &= (1,0,0,\ldots,0,0),\\
		\mathbf{e}_2 &= (0,1,0,\ldots,0,0),\\
		&\mspace{9mu}\vdots\\
		\mathbf{e}_{n+3} &= (0,0,0,\ldots,0,1).
	\end{split}
\end{equation}
We remark
\begin{equation}
	T_k = T_{k,-(k+1)}\quad (k=1,\ldots,n+2),\quad T_{n+3} = T_{n+2,n+3}
\end{equation}
and that they act on $\theta_j$ $(j=1,\ldots,n+3)$ as follows:
\begin{equation}
	\begin{split}
		T_{k,l}(\theta_j) &= \theta_j+\delta_{jk}+\delta_{jl},\\
		T_{k,-l}(\theta_j) &= \theta_j+\delta_{jk}-\delta_{jl},\\
		T_{-k,-l}(\theta_j) &= \theta_j-\delta_{jk}-\delta_{jl}.
	\end{split}
\end{equation}
Then the action of them on the Hamiltonians $H_i$ $(i=1,\ldots,n)$ is described
as follows.
\begin{equation}
	\begin{split}
		T_{k,l}(H_i) &= H_i - \frac{\mathrm{tr}A_iR_{k,l}}{(t_i-t_k)(t_i-t_l)}
		+ \frac{\varGamma^i_k}{t_i-t_k} + \frac{\varGamma^{-i}_l}{t_i-t_l}
		+ \sum_{j=1,j\neq i}^{n+2}\frac{\varGamma_{k,l}}{t_i-t_j}\\
		&\mspace{360mu} (i\neq k,l),\\
		T_{k,l}(H_k) &= H_k - \sum_{j=1,j\neq k,l}^{n+2}
		\frac{\mathrm{tr}A_iR_{k,l}}{(t_k-t_j)(t_k-t_l)}
		- \frac{(n-1)(1+\theta_k+\theta_l)}{2\,(n+1)(t_k-t_l)}\\
		&\qquad\qquad
		+ \sum_{j=1,j\neq k,l}^{n+2}\frac{\varGamma^j_k}{t_k-t_j}
		+ \sum_{j=1,j\neq k}^{n+2}\frac{\varGamma_{k,l}}{t_k-t_j},\\
		T_{k,l}(H_l) &= H_l - \sum_{j=1,j\neq k,l}^{n+2}
		\frac{\mathrm{tr}A_iR_{k,l}}{(t_l-t_j)(t_l-t_k)}
		- \frac{(n-1)(1+\theta_k+\theta_l)}{2\,(n+1)(t_l-t_k)}\\
		&\qquad\qquad
		+ \sum_{j=1,j\neq k,l}^{n+2}\frac{\varGamma^{-j}_k}{t_l-t_j}
		+ \sum_{j=1,j\neq l}^{n+2}\frac{\varGamma_{k,l}}{t_l-t_j},
	\end{split}
\end{equation}
where
\begin{equation}
	\begin{array}{ll}
		\ds\varGamma^j_k
		= -\frac{\theta_j}{2} + \frac{1+2\,\theta_k}{2\,(n+2)},&
		\ds\varGamma^{-j}_k
		= \frac{\theta_j}{2} + \frac{1+2\,\theta_k}{2\,(n+2)},\\[12pt]
		\ds\varGamma_{k,l} = -\frac{1+\theta_k+\theta_l}{(n+1)(n+2)},&
		\ds R_{k,l} = \frac{t_k-t_l}{b_kd_l-d_kb_l}
		\left(\begin{array}{@{}c@{}}b_k\\d_k\end{array}\right)
		\left(\begin{array}{@{}cc@{}}-d_l&b_l\end{array}\right),
	\end{array}
\end{equation}
for $k,l=1,\ldots,n+2$ with $k\neq l$.
\begin{equation}
	\begin{split}
		T_{k,-l}(H_i)
		&= H_i - \frac{\mathrm{tr}A_iR_{k,-l}}{(t_i-t_k)(t_i-t_l)}
		+ \frac{\varGamma^i_k}{t_i-t_k} + \frac{\varGamma^{-i}_{-l}}{t_i-t_l}
		+ \sum_{j=1,j\neq i}^{n+2}\frac{\varGamma_{k,-l}}{t_i-t_j}\\
		&\mspace{360mu} (i\neq k,l),\\
		T_{k,-l}(H_k) &= H_k - \sum_{j=1,j\neq k,l}^{n+2}
		\frac{\mathrm{tr}A_iR_{k,-l}}{(t_k-t_j)(t_k-t_l)}
		- \frac{(n-1)(1+\theta_k-\theta_l)}{2\,(n+1)(t_k-t_l)}\\
		&\qquad\qquad
		+ \sum_{j=1,j\neq k,l}^{n+2}\frac{\varGamma^j_k}{t_k-t_j}
		+ \sum_{j=1,j\neq k}^{n+2}\frac{\varGamma_{k,-l}}{t_k-t_j},\\
		T_{k,-l}(H_l) &= H_l - \sum_{j=1,j\neq k,l}^{n+2}
		\frac{\mathrm{tr}A_iR_{k,-l}}{(t_l-t_j)(t_l-t_k)}
		- \frac{(n-1)(1+\theta_k-\theta_l)}{2\,(n+1)(t_l-t_k)}\\
		&\qquad\qquad
		+ \sum_{j=1,j\neq k,l}^{n+2}\frac{\varGamma^{-j}_{-l}}{t_l-t_j}
		+ \sum_{j=1,j\neq l}^{n+2}\frac{\varGamma_{k,-l}}{t_l-t_j},
    \end{split}
\end{equation}
where
\begin{equation}
	\begin{split}
		&\varGamma_{k,-l} = -\frac{1+\theta_k-\theta_l}{(n+1)(n+2)},\quad
		\varGamma^{-j}_{-k}
		= \frac{\theta_j}{2}+\frac{1-2\,\theta_k}{2\,(n+2)},\\
		&R_{k,-l} = \frac{-t_k+t_l}{b_ka_l+d_kb_l}
		\left(\begin{array}{@{}c@{}}b_k\\d_k\end{array}\right)
		\left(\begin{array}{@{}cc@{}}a_l&b_l\end{array}\right),
	\end{split}
\end{equation}
for $k,l=1,\ldots,n+2$ with $k\neq l$.
\begin{equation}
	\begin{split}
		T_{-k,-l}(H_i)
		&= H_i - \frac{\mathrm{tr}A_iR_{-k,-l}}{(t_i-t_k)(t_i-t_l)}
		+ \frac{\varGamma^i_{-k}}{t_i-t_k}
		+ \frac{\varGamma^{-i}_{-l}}{t_i-t_l}
		+ \sum_{j=1,j\neq i}^{n+2}\frac{\varGamma_{-k,-l}}{t_i-t_j}\\
		&\mspace{360mu} (i\neq k,l),\\
		T_{-k,-l}(H_k) &= H_k - \sum_{j=1,j\neq k,l}^{n+2}
		\frac{\mathrm{tr}A_iR_{-k,-l}}{(t_k-t_j)(t_k-t_l)}
		- \frac{(n-1)(1-\theta_k-\theta_l)}{2\,(n+1)(t_k-t_l)}\\
		&\qquad\qquad
		+ \sum_{j=1,j\neq k,l}^{n+2}\frac{\varGamma^j_{-k}}{t_k-t_j}
		+ \sum_{j=1,j\neq k}^{n+2}\frac{\varGamma_{-k,-l}}{t_k-t_j},\\
		T_{-k,-l}(H_l) &= H_l - \sum_{j=1,j\neq k,l}^{n+2}
		\frac{\mathrm{tr}A_iR_{-k,-l}}{(t_l-t_j)(t_l-t_k)}
		- \frac{(n-1)(1-\theta_k-\theta_l)}{2\,(n+1)(t_l-t_k)}\\
		&\qquad\qquad
		+ \sum_{j=1,j\neq k,l}^{n+2}\frac{\varGamma^{-j}_{-l}}{t_l-t_j}
		+ \sum_{j=1,j\neq l}^{n+2}\frac{\varGamma_{-k,-l}}{t_l-t_j},
	\end{split}
\end{equation}
where
\begin{equation}
	\begin{split}
		&\varGamma_{-k,-l} = -\frac{1-\theta_k-\theta_l}{(n+1)(n+2)},\quad
		\varGamma^j_{-k}
		= -\frac{\theta_j}{2}+\frac{1-2\,\theta_k}{2\,(n+2)},\\
		&R_{-k,-l} = \frac{t_k-t_l}{a_kb_l-b_ka_l}
		\left(\begin{array}{@{}c@{}}b_k\\-a_k\end{array}\right)
		\left(\begin{array}{@{}cc@{}}a_l&b_l\end{array}\right),
	\end{split}
\end{equation}
for $k,l=1,\ldots,n+2$ with $k\neq l$.
\begin{equation}
	\begin{split}
		T_{k,n+3}(H_i)
		&= H_i + \frac{1}{t_i-t_k}\left(a_i+b_i\,\frac{d_k}{b_k}\right)
		+ \frac{\varGamma^i_k}{t_i-t_k} + \sum_{j=1,j\neq i}^{n+2}
		\frac{\varGamma_{k,n+3}}{t_i-t_j}\quad (i\neq k),\\
		T_{k,n+3}(H_k) &= H_k + \sum_{j=1,j\neq k}^{n+2}
		\frac{1}{t_k-t_j}\left(a_j+b_j\,\frac{d_k}{b_k}\right)
		+ \sum_{j=1,j\neq k}^{n+2}
		\frac{\varGamma^j_k+\varGamma_{k,n+3}}{t_k-t_j}
	\end{split}
\end{equation}
for $k=1,\ldots,n+2$.
\begin{equation}
	\begin{split}
		T_{k,-(n+3)}(H_i)
		&= H_i + \frac{1}{t_i-t_k}\left(d_i+c_i\,\frac{a_k}{c_k}\right)
		+ \frac{\varGamma^i_k}{t_i-t_k} + \sum_{j=1,j\neq i}^{n+2}
		\frac{\varGamma_{k,-(n+3)}}{t_i-t_j}\\
		&\mspace{360mu} (i\neq k),\\
		T_{k,-(n+3)}(H_k) &= H_k + \sum_{j=1,j\neq k}^{n+2}
		\frac{1}{t_k-t_j}\left(d_j+c_j\,\frac{a_k}{c_k}\right)
		+ \sum_{j=1,j\neq k}^{n+2}
		\frac{\varGamma^j_k+\varGamma_{k,-(n+3)}}{t_k-t_j}
 	\end{split}
\end{equation}
for $k=1,\ldots,n+2$.
\begin{equation}
	\begin{split}
		T_{n+3,-k}(H_i)
		&= H_i + \frac{1}{t_i-t_k}\left(a_i-b_i\,\frac{a_k}{b_k}\right)
		+ \frac{\varGamma^i_{-k}}{t_i-t_k} + \sum_{j=1,j\neq i}^{n+2}
		\frac{\varGamma_{n+3,-k}}{t_i-t_j}\quad (i\neq k),\\
		T_{n+3,-k}(H_k) &= H_k + \sum_{j=1,j\neq k}^{n+2}
		\frac{1}{t_k-t_j}\left(a_j-b_j\,\frac{a_k}{b_k}\right)
		+ \sum_{j=1,j\neq k}^{n+2}
		\frac{\varGamma^j_{-k}+\varGamma_{n+3,-k}}{t_k-t_j}
	\end{split}
\end{equation}
for $k=1,\ldots,n+2$.
\begin{equation}
	\begin{split}
		T_{-k,-(n+3)}(H_i)
		&= H_i + \frac{1}{t_i-t_k}\left(d_i-c_i\,\frac{d_k}{c_k}\right)
		+ \frac{\varGamma^i_{-k}}{t_i-t_k} + \sum_{j=1,j\neq i}^{n+2}
		\frac{\varGamma_{-k,-(n+3)}}{t_i-t_j}\\
		&\mspace{360mu} (i\neq k),\\
		T_{-k,-(n+3)}(H_k) &= H_k + \sum_{j=1,j\neq k}^{n+2}
		\frac{1}{t_k-t_j}\left(d_j-c_j\,\frac{d_k}{c_k}\right)
		+ \sum_{j=1,j\neq k}^{n+2}
		\frac{\varGamma^j_{-k}+\varGamma_{-k,-(n+3)}}{t_k-t_j}
	\end{split}
\end{equation}
for $k=1,\ldots,n+2$.
For the other $\mu\in L$, the action of $T_{\mu}$ on the Hamiltonians, which is
not described in this paper, is similarly obtained from its action on the
dependent variables.

Next we describe the action of the transformations $\sigma_k$
$(k=1,\ldots,n+2)$ and $r_l$ $(l=1,\ldots,n+3)$ on the Hamiltonians.
Since $H_i$ $(i=1,\ldots,n)$ are invariant under the action of each $\sigma_k$
and $r_l$, we obtain
\begin{equation}
	\sigma_kT_{\mu}(H_i) = T_{\sigma_k(\mu)}(H_i),\quad
	r_lT_{\mu}(H_i) = T_{r_l(\mu)}(H_i)\quad (\mu\in L),
\end{equation}
where
\begin{equation}
	\begin{split}
		\sigma_k(\mu) &= (\mu_{\sigma_k(1)},\ldots,\mu_{\sigma_k(n+3)}),\\
		r_l(\mu) &= (\mu_1,\ldots,\mu_{l-1},-\mu_l,\mu_{l+1},\ldots,\mu_{n+3}).
    \end{split}
\end{equation}

\subsection{Toda equations}
In this section, we present the Toda equations for the Schlesinger
transformations $T_k$ $(k=1,\ldots,n+3)$.
Set
\begin{equation}\label{Eq:Ham_JMU}
	\wt{H}_i = H_i - \sum_{j=1,j\neq i}^{n+2}\frac{C_{ij}}{t_i-t_j}
	= \sum_{j=1,j\neq i}^{n+2}\frac{\mathrm{tr}A_iA_j}{t_i-t_j}\quad
	(i=1,\ldots,n).
\end{equation}
Then we have
\begin{thm}[\rm\cite{JM}\bf]\label{Thm:Tau_Quo}
The Hamiltonians $\wt{H}_i$ $(i=1,\ldots,n)$ satisfy the following
equations{\rm:}
\begin{equation}
	\begin{split}
		T_k(\wt{H}_i) + T_k^{-1}(\wt{H}_i) - 2\,\wt{H}_i
		&= \partial_{t_i}\log
		\frac{(G_{k+1}^{-1}G_k)_{22}(G_k^{-1}G_{k+1})_{22}}{(t_k-t_{k+1})^2}\\
		&\mspace{120mu}(k=1,\ldots,n+1),\\
		T_{n+2}(\wt{H}_i) + T_{n+2}^{-1}(\wt{H}_i)
		- 2\,\wt{H}_i
		&= \partial_{t_i}\log\,(G_{n+2})_{22}(G_{n+2}^{-1})_{22},\\
		T_{n+3}(\wt{H}_i) + T_{n+3}^{-1}(\wt{H}_i)
		- 2\,\wt{H}_i
		&= \partial_{t_i}\log\,(G_{n+2})_{12}(G_{n+2}^{-1})_{21},
	\end{split}
\end{equation}
where $(G_j)_{kl}$ stands for the $(k,l)$-component of the $2\times2$ matrix
$G_j$.
\end{thm}
We also obtain the following lemma.
\begin{lem}\label{Lem:Toda}
The Hamiltonians $\wt{H}_i$ $(i=1,\ldots,n)$ satisfy the
following equations{\rm:}
\begin{equation}\label{Eq:Dif_Ham_Toda}
	\begin{split}
		\partial_{t_k}(\wt{H}_{k+1})
		&= \frac{\mathrm{tr}A_kA_{k+1}}{(t_k-t_{k+1})^2}\quad
		(k=1,\ldots,n-1),\\
		\partial_{t_n}\left(\sum_{i=1}^{n}\,(t_i-1)\wt{H}_i\right)
		&=\frac{\mathrm{tr}A_nA_{n+1}}{t_n^2},\\
		(\delta^*+1)\left(\sum_{i=1}^{n}\,t_i\wt{H}_i\right)
		&= -\mathrm{tr}A_{n+1}A_{n+2}
		- \frac{1}{2}\,\sum_{i=1}^{n+2}\sum_{j=1,j\neq i}^{n+2}C_{ij},\\
		(\delta+1)\left(\sum_{i=1}^{n}\,t_i\wt{H}_i\right)
		&= \theta_{n+3}\,d_{n+2} + \theta_{n+2}(\rho+\theta_{n+2})
		- \frac{1}{2}\,\sum_{i=1}^{n+2}\sum_{j=1,j\neq i}^{n+2}C_{ij},
	\end{split}
\end{equation}
where $\partial_i=\partial/\partial t_i$ and
\begin{equation}
	\delta = \sum_{i=1}^{n}\,t_i(t_i-1)\,\partial_{t_i},\quad
	\delta^* = \sum_{i=1}^{n}\,(t_i-1)\,\partial_{t_i}.
\end{equation}
\end{lem}
{\sl Proof}\quad
The first equation of \eqref{Eq:Dif_Ham_Toda} is obtained by a direct
computation.
The second equation of \eqref{Eq:Dif_Ham_Toda} is obtained by using
\begin{equation}
	\sum_{i=1}^{n}\,(t_i-1)\wt{H}_i
	= -\sum_{j=1,j\neq n+1}^{n+2}\frac{\mathrm{tr}A_jA_{n+1}}{t_j}
	+ \sum_{i=1}^{n+2}\sum_{j=1,j\neq i}^{n+2}\mathrm{tr}A_iA_j
\end{equation}
and
\begin{equation}\label{Eq:Lem_Toda_Prf_1}
	\sum_{i=1}^{n+2}\sum_{j=1,j\neq i}^{n+2}\mathrm{tr}A_iA_j
	= -\sum_{i=1}^{n+2}\sum_{j=1,j\neq i}^{n+2}C_{ij}.
\end{equation}
The third equation of \eqref{Eq:Dif_Ham_Toda} is obtained by using
\eqref{Eq:Lem_Toda_Prf_1},
\begin{equation}\label{Eq:Lem_Toda_Prf_2}
	\sum_{i=1}^{n}\,t_i\wt{H}_i
	= \sum_{j=1}^{n+1}\,\frac{\mathrm{tr}A_jA_{n+2}}{t_j-1}
	+ \frac{1}{2}\,\sum_{i=1}^{n+2}\sum_{j=1,j\neq i}^{n+2}\mathrm{tr}A_iA_j
\end{equation}
and
\begin{equation}
	(\delta^*+1)
	\left(\sum_{j=1}^{n+1}\,\frac{\mathrm{tr}A_jA_{n+2}}{t_j-1}\right)
	= -\mathrm{tr}A_{n+1}A_{n+2}.
\end{equation}
The fourth equation of \eqref{Eq:Dif_Ham_Toda} is obtained by using
\eqref{Eq:Lem_Toda_Prf_1}, \eqref{Eq:Lem_Toda_Prf_2} and
\begin{equation}
	(\delta+1)
	\left(\sum_{j=1}^{n+1}\,\frac{\mathrm{tr}A_jA_{n+2}}{t_j-1}\right)
	= \theta_{n+3}\,d_{n+2} + \theta_{n+2}(\rho+\theta_{n+2}).
\end{equation}
\hfill{$\Box$}\\
{}From Theorem \ref{Thm:Tau_Quo}, Lemma \ref{Lem:Toda} and the following
identities:
\begin{equation}
	\begin{split}
		(G_{k+1}^{-1}G_k)_{22}(G_k^{-1}G_{k+1})_{22}
		&= -\frac{\mathrm{tr}A_kA_{k+1}}{\theta_k\theta_{k+1}}\quad
		(k=1,\ldots,n+1),\\
		(G_{n+2})_{22}(G_{n+2}^{-1})_{22} &= \frac{d_{n+2}}{\theta_{n+2}},\\
		(G_{n+2})_{12}(G_{n+2}^{-1})_{21} &= \frac{a_{n+2}}{\theta_{n+2}},
	\end{split}
\end{equation}
we obtain
\begin{equation}\label{Eq:Toda}
	T_k(\wt{H}_i) + T_k^{-1}(\wt{H}_i) - 2\,\wt{H}_i
	= \partial_{t_i}\log X_k\quad(k=1,\ldots,n+3),
\end{equation}
where
\begin{equation}
	\begin{split}
		X_k &= \partial_{t_k}(\wt{H}_{k+1})\quad (k=1,\ldots,n-1),\\
		X_n &= \partial_{t_n}\left(\sum_{i=1}^{n}\,(t_i-1)\wt{H}_i\right),\\
		X_{n+1} &= (\delta^*+1)\left(\sum_{i=1}^{n}t_i\wt{H}_i\right)
		+ \frac{1}{2}\,\sum_{i=1}^{n+2}\sum_{j=1,j\neq i}^{n+2}C_{ij},\\
		X_{n+2} &= (\delta+1)\left(\sum_{i=1}^{n}t_i\wt{H}_i\right)
		+ \frac{1}{2}\,\sum_{i=1}^{n+2}\sum_{j=1,j\neq i}^{n+2}C_{ij}
		- \theta_{n+2}(\rho+\theta_{n+2}),\\
		X_{n+3} &= (\delta+1)\left(\sum_{i=1}^{n}t_i\wt{H}_i\right)
		+ \frac{1}{2}\,\sum_{i=1}^{n+2}\sum_{j=1,j\neq i}^{n+2}C_{ij}
		- \theta_{n+2}(\rho+\theta_{n+2}+\theta_{n+3}).
	\end{split}
\end{equation}

Here we introduce the Hirota derivatives $D_i$ $(i=1,\cdots,n)$ defined by
\begin{equation}
	P(D_1,\cdots,D_n)\,\varphi\cdot\psi
	= P(\partial_{t_1},\cdots,\partial_{t_n})
	\left(\varphi(s+t)\psi(s-t)\right)\bigm|_{t=0},
\end{equation}
where $P(D_1,\cdots,D_n)$ is a polynomial in the derivations $D_i$
$(i=1,\ldots,n)$.
By the definition, we obtain
\begin{equation}
	\begin{split}
		D_i\,\varphi\cdot\psi &= \partial_{t_i}(\varphi)\,\psi
		- \varphi\,\partial_{t_i}(\psi), \\
		D_iD_j\,\varphi\cdot\psi
		&= \partial_{t_i}\partial_{t_j}(\varphi)\,\psi
		- \partial_{t_i}(\varphi)\,\partial_{t_j}(\psi)
		- \partial_{t_j}(\varphi)\,\partial_{t_i}(\psi)
		+ \psi\,\partial_{t_i}\partial_{t_j}(\varphi)
	\end{split}
\end{equation}
and
\begin{equation}
	\begin{split}
		&\partial_{t_i}\log\frac{\varphi}{\psi}
		= \frac{D_i\,\varphi\cdot\psi}{\varphi\cdot\psi}, \\
		&\partial_{t_i}\partial_{t_j}\log\varphi\psi
		= \frac{D_iD_j\,\varphi\cdot\psi}{\varphi\cdot\psi}
		- \frac{D_i\,\varphi\cdot\psi}{\varphi\cdot\psi}
		\frac{D_j\,\varphi\cdot\psi}{\varphi\cdot\psi}.
	\end{split}
\end{equation}

By substituting \eqref{Eq:Ham_JMU} into \eqref{Eq:Toda}, we obtain the Toda and
Toda-like equations expressed in terms of the Hirota derivatives.
\begin{thm}\label{Thm:Eq_Toda}
For the Schlesinger transformations $T_k$ $(k=1,\ldots,n+3)$, we have the
following Toda and Toda-like equations{\rm:}
\begin{equation}
	\begin{split}
		F_kT_k(\tau_0)\,T_k^{-1}(\tau_0) &= D_kD_{k+1}\tau_0\cdot\tau_0
		- \frac{2\,C_{k\,k+1}}{(t_k-t_{k+1})^2}\,\tau_0^2\quad
		(k=1,\cdots,n-1),\\
		F_nT_n(\tau_0)\,T_n^{-1}(\tau_0)
		&= \sum_{i=1}^{n}\,(t_i-1)D_iD_{n}\tau_0\cdot\tau_0
		+ 2\,\partial_{t_n}(\tau_0)\cdot\tau_0
		- \frac{2\,C_{n\,n+1}}{t_n^2}\,\tau_0^2,\\
		F_{n+1}T_{n+1}(\tau_0)\,T_{n+1}^{-1}(\tau_0)
		&= \sum_{i=1}^{n}\sum_{j=1}^{n}\,(t_i-1)t_jD_iD_{j}\tau_0\cdot\tau_0\\
		&\qquad
		+ 2\,\sum_{i=1}^{n}\,(2\,t_i-1)\,\partial_{t_i}(\tau_0)\cdot\tau_0
		+ 2\,C_{n+1\,n+2}\,\tau_0^2,\\
		F_{n+2}T_{n+2}(\tau_0)\,T_{n+2}^{-1}(\tau_0)
		&= \sum_{i=1}^{n}\sum_{j=1}^{n}\,
		t_i(t_i-1)t_jD_iD_{j}\,\tau_0\cdot\tau_0
		+ 2\,\sum_{i=1}^{n}\,t_i^2\,\partial_{t_i}(\tau_0)\cdot\tau_0\\
		&\qquad
		+ 2\left\{\theta_{n+2}(\rho+\theta_{n+2})
		+ \sum_{j=1}^{n+1}\,C_{i\,n+2}\right\}\tau_0^2,\\
		F_{n+3}T_{n+3}(\tau_0)\,T_{n+3}^{-1}(\tau_0)
		&= \sum_{i=1}^{n}\sum_{j=1}^{n}\,t_i(t_i-1)t_jD_iD_{j}\tau_0\cdot\tau_0
		+ 2\,\sum_{i=1}^{n}\,t_i^2\,\partial_{t_i}(\tau_0)\cdot\tau_0\\
		&\qquad
		+ 2\left\{\theta_{n+2}(\rho+\theta_{n+2}+\theta_{n+3})
		+ \sum_{j=1}^{n+1}\,C_{i\,n+2}\right\}\tau_0^2,\\
	\end{split}
\end{equation}
where
\begin{equation}
	\begin{split}
		F_k &= (t_k-t_{k+1})^{-1/2}
		\prod_{j=1,j\neq k}^{n+2}(t_k-t_j)^{-\varGamma^j_k}
		\prod_{j=1,j\neq k+1}^{n+2}(t_{k+1}-t_j)^{-\varGamma^{-j}_{-k+1}}\\
		&\qquad\times\prod_{i=1}^{n+2}\prod_{j=1,j\neq i}^{n+2}
		(t_i-t_j)^{-\varGamma_{k,-(k+1)}/2}\quad (k=1,\ldots,n+1),\\
		F_{n+2} &= \prod_{j=1}^{n+1}\,(t_j-1)^{-\varGamma^j_{n+2}}
		\prod_{i=1}^{n+2}\prod_{j=1,j\neq i}^{n+2}
		(t_i-t_j)^{-\varGamma_{n+2,-(n+3)}/2},\\
		F_{n+3} &= \prod_{j=1}^{n+1}\,(t_j-1)^{-\varGamma^j_{n+2}}
		\prod_{i=1}^{n+2}\prod_{j=1,j\neq i}^{n+2}
		(t_i-t_j)^{-\varGamma_{n+2,n+3}/2}.
	\end{split}
\end{equation}
\end{thm}
We note that the Toda equation for $T_{n+1}$ is equivalent to the equation
given in \cite{TSU2}.

\subsection{Hirota-Miwa equations}
In the following, we set
\begin{equation}
	\tau_{k,l} = T_{k,l}(\tau_0),\quad \tau_{k,-l}
	= T_{k,-l}(\tau_0)\quad (k,l=1,\ldots,n+3,\,k\neq l).
\end{equation}
We first present the Hirota-Miwa equation for the following six
$\tau$-functions:
\begin{equation*}
	\tau_{n+2,n+3},\quad \tau_{n+1,n+2},\quad \tau_{n+2,-(n+1)},\quad
	\tau_{n+1,n+3},\quad \tau_{n+3,-(n+1)},\quad \tau_0.
\end{equation*}
The action of transformations $T_{n+1,n+2}$, $T_{n+3,-(n+1)}$ and $T_{n+2,n+3}$
on the Hamiltonians $H_i$ $(i=1,\ldots,n)$ is described as follows:
\begin{equation}\label{Eq:Sch_Trf_Ham_HM}
	\begin{split}
		T_{n+1,n+2}(H_i) &= H_i - \frac{\mathrm{tr}A_iR_{n+1,n+2}}{t_i(t_i-1)}
		+ \frac{\varGamma^i_{n+1}}{t_i} + \frac{\varGamma^{-i}_{n+2}}{t_i-1}
		+ \sum_{j=1,j\neq i}^{n+2}\frac{\varGamma_{n+1,n+2}}{t_i-t_j},\\
		T_{n+3,-(n+1)}(H_i)
		&= H_i + \frac{1}{t_i}\left(a_i-b_i\,\frac{a_{n+1}}{b_{n+1}}\right)
		+ \frac{\varGamma^i_{-(n+1)}}{t_i}
		+ \sum_{j=1,j\neq i}^{n+2}\frac{\varGamma_{n+3,-(n+1)}}{t_i-t_j},\\
		T_{n+2,n+3}(H_i)
		&= H_i + \frac{1}{t_i-1}\left(a_i+b_i\,\frac{d_{n+2}}{b_{n+2}}\right)
		+ \frac{\varGamma^i_{n+2}}{t_i-1}
		+ \sum_{j=1,j\neq i}^{n+2}\frac{\varGamma_{n+2,n+3}}{t_i-t_j}.
	\end{split}
\end{equation}
{}From \eqref{Eq:Sch_Trf_Ham_HM} and
\begin{equation}\label{Sys:Pfaff_Tau}
	\begin{split}
		d\log\tau_{k,l} = \sum_{i=1}^{n}\,T_{k,l}(H_i),\quad
		d\log\tau_{k,-l} = \sum_{i=1}^{n}\,T_{k,-l}(H_i)\\
		(k,l=1,\ldots,n+3,\,k\neq l),
	\end{split}
\end{equation}
we obtain
\begin{equation}\label{Eq:Hir_Miw}
	\begin{split}
		\frac{\tau_{n+1,n+2}\,\tau_{n+3,-(n+1)}}{\tau_0\,\tau_{n+2,n+3}}
		&= \left(d_{n+1}-b_{n+1}\frac{d_{n+2}}{b_{n+2}}\right)\\
		&\qquad\times\prod_{i=1}^{n}\,t_i^{1/(n+1)}\,\prod_{i=1}^{n+2}
		\prod_{j=1,j\neq i}^{n+2}(t_i-t_j)^{-1/\{2(n+1)(n+2)\}}.
	\end{split}
\end{equation}
Hence the Hirota-Miwa-equation
\begin{equation}
	\begin{split}
		&\tau_{n+1,n+2}\,\tau_{n+3,-(n+1)}-\tau_{n+1,n+3}\,\tau_{n+2,-(n+1)}\\
		&\qquad = \theta_{n+1}\,\prod_{i=1}^{n}\,t_i^{1/(n+1)}
		\prod_{i=1}^{n+2}\prod_{j=1,j\neq i}^{n+2}
		(t_i-t_j)^{-1/\{2(n+1)(n+2)\}}\,\tau_0\tau_{n+2,n+3}
	\end{split}
\end{equation}
is obtained by the action of the transformation $r_{n+1}$ on the both sides of
\eqref{Eq:Hir_Miw}.

For the other indexies $i,j,k=1,\ldots,n+3$ with $i,j,k$ mutually distinct,
the Hirota-Miwa equations are obtained in a similar way.
\begin{thm}
For any distinct $i,j,k=1,\ldots,n+3$, we have the following Hirota-Miwa
equations{\rm:}
\begin{equation}
	F^{ij}_k\tau_0\,\tau_{i,j}
	= \tau_{i,k}\,\tau_{j,-k} - \tau_{j,k}\,\tau_{i,-k},
\end{equation}
where
\begin{equation}
	\begin{split}
		F^{ij}_k &= \theta_k\,(t_i-t_j)^{1/2}(t_i-t_k)^{-1/2}(t_j-t_k)^{-1/2}\\
		&\qquad\times\prod_{l=1,l\neq k}^{n+2}(t_k-t_l)^{1/(n+1)}
		\prod_{l_1=1}^{n+2}\prod_{l_2=1,l_2\neq l_1}^{n+2}
		(t_{l_1}-t_{l_2})^{-1/\{2(n+1)(n+2)\}},\\
		F^{i,n+3}_j &= \theta_j\prod_{k=1,k\neq j}^{n+2}(t_j-t_k)^{1/(n+1)}
		\prod_{k=1}^{n+2}\prod_{l=1,l\neq k}^{n+2}
		(t_k-t_l)^{-1/\{2(n+1)(n+2)\}},\\
		F^{ij}_{n+3} &= \theta_{n+3}\,(t_i-t_j)^{1/2}\,\prod_{k=1}^{n+2}
		\prod_{l=1,l\neq k}^{n+2}(t_k-t_l)^{-1/\{2(n+1)(n+2)\}}.
	\end{split}
\end{equation}
\end{thm}

\subsection{Bilinear differential equations}
In this section, we present the bilinear differential equations for the
$\tau$-functions $\tau_0$ and $\tau_1=\tau_{n+1,n+2}$.
We set
\begin{equation}
	\wh{H}_i = \sum_{j=1,j\neq i}^{n+2}\frac{t_i(t_i-1)}{t_i-t_j}
	\left(\mathrm{tr}A_iA_j-\frac{1}{2}\,\theta_i\theta_j\right)\quad
	(i=1,\ldots,n)
\end{equation}
and
\begin{equation}
	\wh{H}^*_i = T_{n+1,n+2}(\wh{H}_i)
	= \wh{H}_i - \mathrm{tr}A_iR_{n+1,n+2} + \frac{\theta_i}{2}\quad
	(i=1,\ldots,n).
\end{equation}
Denoting $\wh{R}=R_{n+1,n+2}$, we have
\begin{equation}
	\partial_{t_i}(\wh{R}) = \frac{\wh{R}A_i(\wh{R}-I_2)}{t_i-1}
	- \frac{(\wh{R}-I_2)A_i\wh{R}}{t_i}\quad (i=1,\ldots,n).
\end{equation}
It follows that
\begin{equation}\label{Eq:Der_Ham_Bil}
	\begin{split}
		\delta_i(\wh{H}_i) &= \sum_{j=1,j\neq i}^{n+2}
		\frac{t_i(t_i-1)(t_i^2-2\,t_it_j+t_j)}{(t_i-t_j)^2}
		\left(\mathrm{tr}A_iA_j-\frac{1}{2}\,\theta_i\theta_j\right),\\
		\delta_j(\wh{H}_i) &= \frac{t_i(t_i-1)t_j(t_j-1)}{(t_i-t_j)^2}
		\left(\mathrm{tr}A_iA_j-\frac{1}{2}\,\theta_i\theta_j\right)\quad
		(j=1,\ldots,n,\,j\neq i),\\
		\delta_i(\wh{H}_i-\wh{H}^*_i) &= \mathrm{tr}A_i(\wh{R}-I_2)A_i\wh{R}
		- \sum_{j=1,j\neq i}^{n+2}\frac{t_i(t_i-1)}{t_i-t_j}\,
		\mathrm{tr}\,[A_i,A_j]\wh{R},\\
		\delta_j(\wh{H}_i-\wh{H}^*_i)
		&= t_j\,\mathrm{tr}A_i\wh{R}A_j(\wh{R}-I_2)
		- (t_j-1)\,\mathrm{tr}A_i(\wh{R}-I_2)A_j\wh{R}\\
		&\qquad
		- \frac{t_j(t_j-1)}{t_i-t_j}\,\mathrm{tr}\,[A_i,A_j]\wh{R}\quad
		(j=1,\ldots,n,\,j\neq i),
	\end{split}
\end{equation}
where $\delta_i=t_i(t_i-1)\,\partial_i$, for each $i=1,\ldots,n$.
By using \eqref{Eq:Der_Ham_Bil}, we obtain
\begin{equation}\label{Eq:Bil_LHS}
	\begin{split}
		&\sum_{j=1}^{n}\,\frac{2}{2\,t_it_j-t_i-t_j}
		\left\{\delta_j(\wh{H}_i+\wh{H}^*_i)
		+ (\wh{H}_i-\wh{H}^*_i)(\wh{H}_j-\wh{H}^*_j)\right\}\\
		&\qquad = -\frac{\mathrm{tr}A_i(\wh{R}-I_2)A_i\wh{R}}{t_i(t_i-1)}
		+ \frac{1}{t_i(t_i-1)}
		\left(\mathrm{tr}A_i\wh{R}-\frac{\theta_i}{2}\right)^2\\
		&\qquad\qquad + \sum_{j=1,j\neq i}^{n}\,\frac{2}{2\,t_it_j-t_i-t_j}
		\biggl\{\left(\mathrm{tr}A_i\wh{R}-\frac{\theta_i}{2}\right)
		\left(\mathrm{tr}A_j\wh{R}-\frac{\theta_j}{2}\right)\\
		&\qquad\qquad\qquad
		+ (t_j-1)\,\mathrm{tr}A_i(\wh{R}-I_2)A_j\wh{R}
		- t_j\,\mathrm{tr}A_i\wh{R}A_j(\wh{R}-I_2)\biggr\}\\
		&\qquad\qquad + \sum_{j=1,j\neq i}^{n+2}\,\frac{1}{2\,t_it_j-t_i-t_j}
		\left\{(2\,t_j-1)\,\mathrm{tr}\,[A_i,A_j]\wh{R}
		- \mathrm{tr}A_iA_j + \frac{1}{2}\,\theta_i\theta_j\right\}\\
		&\qquad\qquad + \sum_{j=1,j\neq i}^{n+2}\,\frac{2\,t_i-1}{t_i-t_j}
		\left(\mathrm{tr}A_iA_j-\frac{1}{2}\,\theta_i\theta_j\right)\quad
		(i=1,\ldots,n).
	\end{split}
\end{equation}
On the other hand, we obtain
\begin{equation}\label{Eq:Bil_RHS_1}
	\begin{split}
		\mathrm{tr}A_i(\wh{R}-I_2)A_j\wh{R}
		&= \left(\mathrm{tr}A_i\wh{R}-\frac{\theta_i}{2}\right)
		\left(\mathrm{tr}A_j\wh{R}-\frac{\theta_j}{2}\right)
		- \frac{1}{2}\,\mathrm{tr}\,[A_i,A_j]\wh{R}\\
		&\qquad
		- \frac{1}{2}\,\mathrm{tr}A_iA_j + \frac{1}{4}\,\theta_i\theta_j\quad
		(j=1,\ldots,n,\,j\neq i),\\
		\mathrm{tr}A_i\wh{R}A_j(\wh{R}-I_2)
		&= \left(\mathrm{tr}A_i\wh{R}-\frac{\theta_i}{2}\right)
		\left(\mathrm{tr}A_j\wh{R}-\frac{\theta_j}{2}\right)
		+ \frac{1}{2}\,\mathrm{tr}\,[A_i,A_j]\wh{R}\\
		&\qquad
		- \frac{1}{2}\,\mathrm{tr}A_iA_j + \frac{1}{4}\,\theta_i\theta_j\quad
		(j=1,\ldots,n,\,j\neq i),\\
		\mathrm{tr}A_i(\wh{R}-I_2)A_i\wh{R}
		&= \left(\mathrm{tr}A_i\wh{R}-\frac{\theta_i}{2}\right)^2
		- \frac{\theta_i^2}{4\,t_i(t_i-1)}
	\end{split}
\end{equation}
and
\begin{equation}\label{Eq:Bil_RHS_2}
	\begin{split}
		\mathrm{tr}[A_i,A_{n+1}]\wh{R} + \mathrm{tr}A_iA_{n+1}
		&= \theta_{n+1}\,\mathrm{tr}A_i\wh{R},\\
		\mathrm{tr}[A_i,A_{n+2}]\wh{R} - \mathrm{tr}A_iA_{n+2}
		&= \theta_{n+2}\,\mathrm{tr}A_i(\wh{R}-I_2)
	\end{split}
\end{equation}
by direct computations for each $i=1,\ldots,n$.
{}From \eqref{Eq:Bil_LHS}, \eqref{Eq:Bil_RHS_1} and \eqref{Eq:Bil_RHS_2}, the
following differential equations are obtained:
\begin{equation}\label{Eq:Bil}
	\begin{split}
		&\sum_{j=1}^{n}\,\frac{2}{2\,t_it_j-t_i-t_j}
		\left\{\delta_j(\wh{H}_i+\wh{H}^*_i)
		+ (\wh{H}_i-\wh{H}^*_i)(\wh{H}_j-\wh{H}^*_j)\right\}\\
		&\qquad = \left(\frac{\theta_{n+1}}{t_i}
		+ \frac{\theta_{n+2}}{t_i-1}\right)(\wh{H}_i-\wh{H}^*_i)
		+ \frac{2\,t_i-1}{t_i(t_i-1)}\,\wh{H}_i
		+ \frac{\theta_i^2}{4\,t_i(t_i-1)}\\
		&\mspace{360mu} (i=1,\ldots,n).
	\end{split}
\end{equation}
By substituting
\begin{equation}
	\wh{H}_i = \delta_i\log\tau_0 + \wh{C}_i,\quad
	\wh{H}^*_i = \delta_i\log\tau_1 + \wh{C}^*_i,
\end{equation}
where
\begin{equation}
	\wh{C}_i = \sum_{j=1,j\neq i}^{n+2}\frac{t_i(t_i-1)}{t_i-t_j}
	\left(C_{ij}+\frac{1}{2}\,\theta_i\theta_j\right),\quad
	\wh{C}^*_i = T_{n+1,n+2}(\wh{C}_i)
\end{equation}
into \eqref{Eq:Bil}, we obtain the bilinear differential equations for the
$\tau$-functions $\tau_0$ and $\tau_1$.
\begin{thm}
The $\tau$-functions $\tau_0$ and $\tau_1$ satisfy the following bilinear
differential equations{\rm:}
\begin{equation}
	\begin{split}
		&\sum_{j=1}^{n}\,\frac{2}{2\,t_it_j-t_i-t_j}
		\left\{D^{*}_iD^{*}_j\,\tau_0\cdot\tau_1
		+ F^i_jD^{*}_j\,\tau_0\cdot\tau_1\right\}
		+ F^{i,0}D^{*}_i\,\tau_0\cdot\tau_1\\
		&\qquad - \frac{2\,t_i-1}{t_i(t_i-1)}\,\delta_i(\tau_0)\cdot\tau_1
		+ F^{i,1}\tau_0\cdot\tau_1 = 0\quad (i=1,\ldots,n),
	\end{split}
\end{equation}
where
\begin{equation}
	\begin{split}
		F^i_j &= \wh{C}_i - \wh{C}^*_i,\\
		F^{i,0}
		&= \sum_{j=1}^{n}\,\frac{2\,(\wh{C}_i-\wh{C}^*_i)}{2\,t_it_j-t_i-t_j}
		- \frac{\theta_{n+1}}{t_i} - \frac{\theta_{n+2}}{t_i-1},\\
		F^{i,1} &= \sum_{j=1}^{n}\,\frac{2}{2\,t_it_j-t_i-t_j}
		\left\{\delta_j(\wh{C}_i+\wh{C}^*_i)
		+ (\wh{C}_i-\wh{C}^*_i)(\wh{C}_j-\wh{C}^*_j)\right\}\\
		&\qquad - \left(\frac{\theta_{n+1}}{t_i}
		+ \frac{\theta_{n+2}}{t_i-1}\right)(\wh{C}_i-\wh{C}^*_i)
		- \frac{2\,t_i-1}{t_i(t_i-1)}\,\wh{C}_i
		- \frac{\theta_i^2}{4\,t_i(t_i-1)}
	\end{split}
\end{equation}
and $D^*_i$ stands for the Hirota derivative with respect to the derivation
$\delta_i$.
\end{thm}

\section{Garnier system}\label{Sec:Gar}
We consider rational functions in $a_j$, $b_j$, $c_j$, $d_j$ $(j=1,\ldots,n+2)$
defined as
\begin{equation}\label{Eq:Sch_to_Gar}
	\begin{array}{ll}
		\ds q_i = \frac{t_ib_i}{b_{\infty}}& (i=1,\ldots,n),\\[12pt]
		\ds p_i = \frac{b_{\infty}}{t_i}\left\{\frac{a_i}{b_i}
		+ (t_i-1)\,\frac{a_{n+1}}{b_{n+1}}
		- t_i\,\frac{a_{n+2}}{b_{n+2}}\right\}& (i=1,\ldots,n),\\[12pt]
		\ds x_i = \frac{t_i}{t_i-1}& (i=1,\ldots,n),
	\end{array}
\end{equation}
where $b_{\infty}=\sum_{j=1}^{n+2}t_jb_j$.
Let $\{\,,\,\}$ be the Poisson bracket defined by
\begin{equation}
	\{\varphi,\psi\}
	= \sum_{j=1}^{n}\left(\frac{\partial\varphi}{\partial p_j}
	\frac{\partial\psi}{\partial q_j}
	- \frac{\partial\varphi}{\partial q_j}
	\frac{\partial\psi}{\partial p_j}\right).
\end{equation}
Also let $\bar{d}$ be an exterior differentiation with respect to
$x_1,\ldots,x_n$.
Then we have
\begin{prop}[\rm\cite{IKSY}\bf]
The independent and dependent variables $q_i$, $p_i$, $x_i$ $(i=1,\ldots,n)$
defined by \eqref{Eq:Sch_to_Gar} satisfy the Garnier system
\begin{equation}\label{Sys:Garnier}
	\bar{d}q_i = \sum_{j=1}^{n}\,\{\bar{H}_j,q_i\}\,dx_j,\quad
	\bar{d}p_i = \sum_{j=1}^{n}\,\{\bar{H}_j,p_i\}\,dx_j
\end{equation}
with the Hamiltonians
\begin{equation}\label{Eq:Sch2Gar_Tau}
	-(x_i-1)^2\,\bar{H}_i = T_{n+3,-(n+1)}(H_i)\quad (i=1,\ldots,n).
\end{equation}
\end{prop}

Here we remark
\begin{equation}
	\bar{H}_i = K_i+\sum_{j=1,j\neq i}^{n+2}\frac{\bar{C}_{ij}}{x_i-x_j}\quad
	(i=1,\ldots,n),
\end{equation}
where
\begin{equation}
	\begin{split}
		\bar{C}_{ij} &= T_{n+3,-(n+1)}(C_{ij})+\theta_i\theta_j\quad
		(j=1,\ldots,n),\\
		\bar{C}_{i\,n+1}
		&= T_{n+3,-(n+1)}(C_{i\,n+1})+\theta_i(\theta_{n+1}-1),\\
		\bar{C}_{i\,n+2} &= -\sum_{j=1,j\neq i}^{n+2}T_{n+3,-(n+1)}(C_{ij})
		+ \theta_i(\theta_i+\theta_{n+3}+2\rho+1)
	\end{split}
\end{equation}
and $K_i$ is given by \eqref{Eq:Ham_Gar}.

In this section, we show that the Garnier system has affine Weyl group symmetry
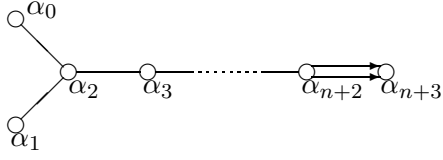
of type $B_{n+3}^{(1)}$. We also show that the $\tau$-functions for the Garnier
system, formulated on the root lattice $Q(C_{n+3})$, satisfy Toda equations,
Hirota-Miwa equations and bilinear differential equations.

\subsection{Affine Weyl group symmetries}\label{SubSec:Sym_Gar}
The transformations $\sigma_k$, $r_l$ and $T_{\mu}$ given in Section
\ref{Sec:Sch} can be lifted to the birational canonical transformations of the
variables $q_i$, $p_i$, $x_i$ $(i=1,\ldots,n)$ which is already known in
\cite{TSU1,TSU2}.
In this section, we formulate the action of those transformations as
realization of affine Weyl group.

Denote the parameter by
\begin{equation}\label{Eq:Sch_to_Gar_Par}
	\begin{split}
		&\vep_1 = \theta_{n+1},\quad \vep_2 = \theta_{n+2},\quad
		\vep_3 = \theta_{n+3}+1,\\
		&\vep_j = \theta_{j-3}\quad (j=4,\ldots,n+3).
	\end{split}
\end{equation}
Then the group of symmetries for the Garnier system is generated by the
transformations $s_k$ $(k=0,1,\ldots,n+3)$ which act on $\vep_j$
$(j=1,\ldots,n+3)$ as follows:
\begin{equation}
	\begin{split}
		&s_0(\vep_1) = 1-\vep_2,\quad s_0(\vep_2) = 1-\vep_1,\quad
		s_0(\vep_j) = \vep_j\quad (j\neq1,2),\\
		&s_k(\vep_j) = \vep_{\sigma_k(j)}\quad (k=1,\ldots,n+2),\\
		&s_{n+3}(\vep_j) = (-1)^{\delta_{jn+3}}\vep_j\quad (j\neq n+3).
	\end{split}
\end{equation}
We describe the action of $s_k$ on the variables $q_i$, $p_i$, $x_i$
$(i=1,\ldots,n)$.
\begin{equation}
	s_0(q_j) = \frac{p_j(q_jp_j-\vep_{j+3})}{Q_1(Q_1+\vep_3)},\quad
	s_0(q_jp_j) = \vep_{j+3} - q_jp_j,\quad s_0(x_i) = \frac{1}{x_i},
\end{equation}
where
\begin{equation}
	Q_1 = \sum_{l=1}^{n}\,q_lp_l
	+ \frac{1}{2}\left(1-\sum_{l=1}^{n+3}\,\vep_l\right),
\end{equation}
for $k=0$.
\begin{equation}
	s_1(q_j) = \frac{q_j}{x_j},\quad s_1(p_j) = x_jp_j,\quad
	s_1(x_i) = \frac{1}{x_i}
\end{equation}
for $k=1$.
\begin{equation}
	s_2(q_j) = \frac{q_j}{Q_2},\quad s_2(p_j) = (p_j-Q_1)Q_2,\quad
	s_2(x_i) = \frac{x_i}{x_i-1},
\end{equation}
where
\begin{equation}
	Q_2 = \sum_{j=1}^{n}\,q_j - 1,
\end{equation}
for $k=2$.
\begin{equation}
	\begin{array}{lll}
		\ds s_3(q_1) = \frac{1}{q_1},&
		\ds s_3(q_j) = -\frac{q_j}{q_1}& (j\neq 1),\\[8pt]
		\ds s_3(p_1) = -q_1Q_1,& \ds s_3(p_j) = -q_1p_j& (j\neq 1),\\[8pt]
		\ds s_3(x_1) = \frac{1}{x_1},&
		\ds s_n(x_i) = \frac{x_i}{x_1}& (i\neq 1)
	\end{array}
\end{equation}
for $k=3$.
\begin{equation}
	s_k(q_j) = q_{\sigma_{k-3}(j)},\quad p_k(q_j) = p_{\sigma_{k-3}(j)},\quad
	s_k(x_j) = x_{\sigma_{k-3}(j)}
\end{equation}
for $k=4,\ldots,n+2$.
\begin{equation}
	\begin{split}
		&s_{n+3}(q_j) = q_j,\\
		&s_{n+3}(p_n) = p_n - \frac{\vep_{n+3}}{q_n},\quad
		s_{n+3}(p_j) = p_j\quad (j\neq n),\\
		&s_{n+3}(x_i) = x_i
	\end{split}
\end{equation}
for $k={n+3}$.
The group generated by these $s_k$ is isomorphic to affine Weyl group
$W(B^{(1)}_{n+3})$.
\begin{thm}
The birational canonical transformations $s_k$ $(k=0,\ldots,n+3)$ satisfy the
fundamental relations for the generators of $W(B^{(1)}_{n+3})$
\begin{equation}
	\begin{array}{lll}
		s_k^2=1& (k=0,\ldots,n+3),\\[4pt]
		(s_ks_l)^2=1& (k,l\neq0,1,2,\,|k-l|>1),\\[4pt]
		(s_ks_{k+1})^3=1& (k=1,\ldots,n+1),\\[4pt]
		(s_0s_1)^2=1,& (s_0s_2)^3=1,\qquad (s_{n+2}s_{n+3})^4=1.
	\end{array}
\end{equation}
\end{thm}
The simple affine roots of $B^{(1)}_{n+3}$ is given as\DynB
\begin{equation}
	\begin{array}{ll}
		\alpha_0 = 1-\vep_1-\vep_2,\\[4pt]
		\alpha_j = \vep_j-\vep_{j+1}& (j=1,\ldots,n+2),\\[4pt]
		\alpha_{n+3} = \vep_{n+3}
	\end{array}
\end{equation}
and the action of $s_k$ on $\alpha_j$ $(j=0,1,\ldots,n+3)$ is described as
follows.
\begin{equation}
	s_0(\alpha_0)=-\alpha_0,\quad s_0(\alpha_2)=\alpha_0+\alpha_2,\quad
	s_0(\alpha_j)=\alpha_j\quad (j\neq0,2)
\end{equation}
for $k=0$.
\begin{equation}
	s_1(\alpha_1)=-\alpha_1,\quad s_1(\alpha_2)=\alpha_1+\alpha_2,\quad
	s_1(\alpha_j)=\alpha_j\quad (j\neq0,1)
\end{equation}
for $k=1$.
\begin{equation}
	\begin{array}{ll}
		s_2(\alpha_2)=-\alpha_2,\\[4pt]
		s_2(\alpha_j)=\alpha_j+\alpha_2& (j=0,1,3),\\[4pt]
		s_2(\alpha_j)=\alpha_j& (j\neq0,1,2,3)
	\end{array}
\end{equation}
for $k=2$.
\begin{equation}
	\begin{split}
		&s_k(\alpha_k) = -\alpha_k,\quad
		s_k(\alpha_{k+1}) = \alpha_{k+1}+\alpha_k,\quad
		s_k(\alpha_{k-1}) = \alpha_{k-1}+\alpha_k,\\
		&s_k(\alpha_j)=\alpha_j\quad (j\neq k,k+1,k-1)
	\end{split}
\end{equation}
for $k=3,\ldots,n+2$.
\begin{equation}
	\begin{split}
		&s_{n+3}(\alpha_{n+3})=-\alpha_{n+3},\quad
		s_{n+3}(\alpha_{n+2})=\alpha_{n+2}+2\,\alpha_{n+3},\\
		&s_{n+3}(\alpha_j)=\alpha_j\quad (j\neq n+2,n+3)
	\end{split}
\end{equation}
for $k=n+3$.

\begin{rem}
The group generated by the transformations $s_1,\ldots,s_{n+2}$ is isomorphic
to the symmetric group $\mathfrak{S}_{n+3}$ {\rm\cite{IKSY}}.
Furthermore, the group generated by $s_1,\ldots,s_{n+3}$ is isomorphic to
$W(B_{n+3})${\rm;} e.g. {\rm\cite{OKM1}}.
\end{rem}
\begin{rem}
In the only case $n=1$, there is the following birational canonical
transformation{\rm:}
\begin{equation}
	\begin{split}
		&s^*_0(q) = q - \frac{\vep_4}{p},\quad s^*_0(p) = p,\quad
		s^*_0(t) = t,\\
		&s^*_0(\vep_j)
		= \vep_j + \frac{1}{2}\,(1-\vep_1-\vep_2-\vep_3-\vep_4)\quad
		(j=1,\ldots,4).
	\end{split}
\end{equation}
The transformation $s_0$ is generated by a composition of $s^*_0$ and
$s_1,\ldots,s_4$.
\end{rem}
But $s^*_0$ cannot be generated by a composition of $s_0,\,s_1,\ldots,s_4$.
It follows that the group of symmetries for the Garnier system in 1-variable
contains affine Weyl group $W(B_4^{(1)})$.
Actually, it is known that $P_{VI}$ has affine Weyl group symmetry of type
$F^{(1)}_4$.
The simple affine roots of $F_4^{(1)}$ is given by\DynF
\begin{equation}
	\begin{array}{ll}
		\alpha_0 = \vep_1 - \vep_2,& \alpha_1 = \vep_2 - \vep_3,\quad
		\alpha_2 = \vep_3 - \vep_4,\\[4pt]
		\alpha_3 = \vep_4,&
		\ds\alpha_4 = \frac{1}{2}\,(1-\vep_1-\vep_2-\vep_3-\vep_4)
	\end{array}
\end{equation}
and $s^*_0$, $s_1,\ldots,s_4$ act on $\alpha_j$ $(j=0,1,\ldots,4)$ as follows:
\begin{equation}
	\begin{array}{llll}
		s^*_0(\alpha_4) = -\alpha_4,& s^*_0(\alpha_3) = \alpha_3+\alpha_4,&
		s^*_0(\alpha_j) = \alpha_j& (j\neq3,4),\\[4pt]
		s_1(\alpha_0) = -\alpha_0,& s_1(\alpha_1)=\alpha_1+\alpha_0,&
		s_1(\alpha_j) = \alpha_j& (j\neq0,1),\\[4pt]
		s_2(\alpha_1) = -\alpha_1,& s_2(\alpha_i)=\alpha_i+\alpha_1,&
		s_2(\alpha_j) = \alpha_j& (i=0,2,\,j=3,4),\\[4pt]
		s_3(\alpha_2) = -\alpha_2,& s_3(\alpha_i)=\alpha_i+\alpha_2,&
		s_3(\alpha_j) = \alpha_j& (i=1,3,\,j=1,4),\\[4pt]
		s_4(\alpha_3) = -\alpha_3,& s_4(\alpha_2)=\alpha_2+2\,\alpha_3,&
		s_4(\alpha_4) = \alpha_4+\alpha_3,\\[4pt]
		s_4(\alpha_j) = \alpha_j& & & (j=1,2).
	\end{array}
\end{equation}

\subsection{$\tau$-Functions}
For each solution of the Garnier system, we introduce the $\tau$-functions
$\bar{\tau}_{\mu}$ $(\mu\in L)$ satisfying the Pfaffian systems
\begin{equation}
	\bar{d}\log\bar{\tau}_{\mu} = \sum_{i=1}^{n}\,T_{\mu}(\bar{H}_i)\,dx_i.
\end{equation}
Each $\bar{\tau}_{\mu}$ is determined up to multiplicative constants.
{}From \eqref{Eq:Sch2Gar_Tau}, we can identify these $\bar{\tau}_{\mu}$ with
the $\tau$-functions for the Schlesinger system by
\begin{equation}\label{Eq:Sch_to_Gar_Tau}
	\bar{\tau}_0 = \tau_{n+3,-(n+1)}.
\end{equation}
Hence we can apply the properties of the $\tau$-functions $\tau_{\mu}$ system to
the Garnier system.
For each $\mu\in L$, the action of the birational canonical transformations
$s_k$ on $\bar{\tau}_{\mu}$ is defined by
\begin{equation}
	s_k(\bar{\tau}_{\mu}) = \bar{\tau}_{s_k(\mu)}\quad
	(k=0,1,\ldots,n+3),
\end{equation}
where
\begin{equation}
	\begin{split}
		s_0(\mu) &= (1-\mu_2,1-\mu_1,\mu_3,\ldots,\mu_{n+3}),\\
		s_k(\mu) &= (\mu_{(k,k+1)1},\ldots,\mu_{(k,k+1)(n+3)})\quad
		(k=1,\ldots,n+2),\\
		s_{n+3}(\mu) &= (\mu_1,\ldots,\mu_{n+2},-\mu_{n+3})
	\end{split}
\end{equation}
and $(k,k+1)$ stands for the adjacent transpositions.
We also obtain bilinear relations which are satisfied by $\bar{\tau}_{\mu}$
formulated on the root lattice $Q(C_{n+3})$.
\begin{thm}
The $\tau$-functions $\bar{\tau}_{\mu}$ $(\mu\in L)$ satisfy the Toda
equations, the Hirota-Miwa equations and the bilinear differential equations
given in Section \ref{Sec:Tau_Sch}.
\end{thm}

In the last, we present the following proposition.
\begin{prop}\label{Prop:Tau_to_Gar}
For the $\tau$-functions
\begin{equation*}
	\bar{\tau}_{1,-2} = \bar{\tau}_{\mathbf{e}_1-\mathbf{e}_2},\quad
	\bar{\tau}_{1,3} = \bar{\tau}_{\mathbf{e}_1+\mathbf{e}_3},\quad
	\bar{\tau}_{1,-3} = \bar{\tau}_{\mathbf{e}_1-\mathbf{e}_3}
\end{equation*}
and $\bar{\tau}_0$, the following relations are satisfied{\rm:}
\begin{equation}\label{Eq:Tau_to_Gar}
	\begin{split}
		q_i &= -\frac{1}{\vep_3}\,x_i(x_i-1)\,\frac{\partial}{\partial x_i}
		\log\frac{\bar{\tau}_{1,3}}{\bar{\tau}_{1,-3}}
		+ 2\,\bar{X}_i\quad (i=i,\ldots,n),\\
		q_ip_i &= -x_i\,\frac{\partial}{\partial x_i}
		\log\frac{\bar{\tau}_{1,-2}}{\bar{\tau}_0}
		+ \frac{\bar{\varGamma}^{j+3}_{-1}-x_i\bar{\varGamma}^{j+3}_{-2}
		- (\vep_1-\vep_2)\bar{X}_i}{x_i-1}\quad (i=1,\ldots,n),
	\end{split}
\end{equation}
where
\begin{equation}
	\bar{X}_{i}
	= \sum_{j=1,j\neq i}^{n+2}\frac{x_i(x_j-1)}{(n+1)(n+2)(x_i-x_j)},\quad
	\bar{\varGamma}^i_{-k} = -\frac{\vep_i}{2} + \frac{1-2\,\vep_k}{2\,(n+1)}.
\end{equation}
\end{prop}
{\sl Proof}\quad
By using \eqref{Eq:Sch_to_Gar}, \eqref{Eq:Sch_to_Gar_Par} and
\eqref{Eq:Sch_to_Gar_Tau}, we can rewrite the relations \eqref{Eq:Tau_to_Gar}
into
\begin{equation}\label{Eq:Tau_to_Sch}
	\begin{split}
		q_i &= \frac{t_i}{\theta_{n+3}+1}\,
		\partial_i\log\frac{\tau_{2\,\mathbf{e}_{n+3}}}{\tau_0}
		- \sum_{j=1,j\neq i}^{n+2}\frac{2\,t_i}{(n+1)(n+2)(t_i-t_j)},\\
		q_ip_i &= t_i(t_i-1)\,
		\partial_i\log\frac{\tau_{n+3,-(n+2)}}{\tau_{n+3,-(n+1)}}
		+ (t_i-1)\,\varGamma^{i}_{-(n+1)} - t_i\varGamma^{i}_{-(n+2)}\\
		&\qquad
		+ \sum_{j=1,j\neq i}^{n+2}\frac{t_i(t_i-1)(\theta_{n+1}-\theta_{n+2})}
		{(n+1)(n+2)(t_i-t_j)}\qquad (i=1,\ldots,n),
	\end{split}
\end{equation}
where
\begin{equation}
	\varGamma^i_{-k}
	= -\frac{\theta_i}{2} + \frac{1-2\,\theta_k}{2\,(n+1)}.
\end{equation}
Hence we show the relations \eqref{Eq:Tau_to_Sch} in the following.

We consider the Schlesinger transformations $T_{2\,\mathbf{e}_{n+3}}$ which act
on the parameters as follows:
\begin{equation}
	T_{2\,\mathbf{e}_{n+3}}(\theta_j) = \theta_j+2\,\delta_{j\,n+3}\quad
	(j=1,\ldots,n+3).
\end{equation}
The action of $T_{2\mathbf{e}_{n+3}}$ on the Hamiltonians $H_i$
$(i=1,\ldots,n)$ is described as follows:
\begin{equation}\label{Eq:Sch_Trf_Q}
	T_{2\,\mathbf{e}_{n+3}}(H_i)
	= H_i + (\theta_{n+3}+1)\,\frac{b_i}{b_{\infty}}
	+ \sum_{j=1,j\neq i}^{n+2}\frac{2\,(\theta_{n+3}+1)}{(n+1)(n+2)(t_i-t_j)}.
\end{equation}
{}From \eqref{Sys:Pfaff_Tau} and \eqref{Eq:Sch_Trf_Q}, the first relation of
\eqref{Eq:Tau_to_Sch} is obtained.
The second relation of \eqref{Eq:Tau_to_Sch} is obtained in a similar way.
\hfill{$\Box$}\\

{\bf Acknowledgement}\quad
The auther is grateful to Professors Masatoshi Noumi, Masa-Hiko Saito and
Yasuhiko Yamada for valuable discussions and advices.


\end{document}